\documentclass[sigchi]{acmart}

\AtBeginDocument{%
  \providecommand\BibTeX{{%
    \normalfont B\kern-0.5em{\scshape i\kern-0.25em b}\kern-0.8em\TeX}}}

\copyrightyear{2020}
\acmYear{2020}
\setcopyright{rightsretained}
\acmConference[IUI '20 Workshops]{March 17, 2020}{Cagliari, Italy}

\begin{document}

\title[Snooze-less Notifications]{A Snooze-less User-Aware Notification System for Proactive Conversational Agents}

\author{Yara Rizk}
\email{yara.rizk@ibm.com}
\affiliation{%
  \institution{IBM Research AI}
  \streetaddress{75 Binney Street}
  \city{Cambridge}
  \state{Massachusetts}
  \postcode{02142}
}
\author{Vatche Isahagian}
\email{vatchei@ibm.com}
\affiliation{%
  \institution{IBM Research AI}
  \streetaddress{75 Binney Street}
  \city{Cambridge}
  \state{Massachusetts}
  \postcode{02142}
}
\author{Merve Unuvar}
\email{munuvar@us.ibm.com}
\affiliation{%
  \institution{IBM Research AI}
  \streetaddress{75 Binney Street}
  \city{Cambridge}
  \state{Massachusetts}
  \postcode{02142}
}
\author{Yasaman Khazaeni}
\email{yasaman.khazaeni@us.ibm.com}
\affiliation{%
  \institution{IBM Research AI}
  \streetaddress{75 Binney Street}
  \city{Cambridge}
  \state{Massachusetts}
  \postcode{02142}
}

\renewcommand{\shortauthors}{Rizk et al.}

\begin{abstract}
The ubiquity of smart phones and electronic devices has placed a wealth of information at the fingertips of consumers as well as creators of digital content. This has led to millions of notifications being issued each second from alerts about posted YouTube videos to tweets, emails and personal messages. Adding work related notifications and we can see how quickly the number of notifications increases. Not only does this cause reduced productivity and concentration but has also been shown to cause alert fatigue. This condition makes users desensitized to notifications, causing them to ignore or miss important alerts. Depending on what domain users work in, the cost of missing a notification can vary from a mere inconvenience to life and death. Therefore, in this work, we propose an alert and notification framework that intelligently issues, suppresses and aggregates notifications, based on event severity, user preferences, or schedules, to minimize the need for users to ignore, or snooze their notifications and potentially forget about addressing important ones. Our framework can be deployed as a backend service, but is better suited to be integrated into proactive conversational agents, a field receiving a lot of attention with the digital transformation era, email services, news services and others. However, the main challenge lies in developing the right machine learning algorithms that can learn models from a wide set of users while customizing these models to individual users' preferences. 
\end{abstract}



\maketitle

\sloppy

\section{Introduction}
Due to increased adoption of digital transformation, we are bombarded with hundreds of notifications every day on our smart phones (and other electronic devices) from calendar reminders, Twitter notifications, and Facebook posts. Then, there are work related event notifications including work emails, calendar invites, deadline reminders, etc. 

Handling notifications as soon as they arrive can cause disruption of work, loss of focus on specific tasks, and decrease of overall productivity in the workplace,  especially when notifications arrive at the rate of minutes instead of hours. Checking them in bulk increases the risk of important notifications falling through the cracks. Furthermore, depending on when users prefer to check their messages (mornings versus lunch breaks or evenings), missed deadlines and delayed response to urgent alerts may increase.  

Even worse is experiencing alert fatigue, a problem faced in many domains from business to healthcare \cite{murphy2012notifications}, where too many notifications make individuals desensitized to them. This, in turn, causes them to miss some critical notifications that may have detrimental consequences. For example, in healthcare, critical test results may be missed, leading to the death or sever impairment of patients. In business, significant amounts of capital may be lost or workers may be put in harms way as a result of missed alerts. 

In this work, we propose an alerting framework that can be embedded within proactive conversational agents or deployed as a backend service, to issue, suppress and aggregate notifications. This system attempts to reduce alert fatigue through selective issuance and suppression of notifications. A learning algorithm models the behavior of the observed system (e.g. estimates the number of expected alerts) and user preference to determine which alerts to issue, suppress or aggregate and the means of alert notification. User preference can be learned by observing user behavior and fine tuned with explicit feedback from the user through natural language conversation with the proactive chatbot. In addition to reducing alert fatigue and minimizing the risk of missing important notification, this framework enhances worker productivity and concentration by minimizing distractions. The main set of research questions that must be addressed to enable such an alerting framework has to do with \textit{how do we learn from user data the effective user-specific rules to issue, suppress and aggregate notifications}. 

Next, Section \ref{sec:litRev} briefly surveys related work. Section \ref{sec:basics} introduces the background and notation. Section \ref{sec:method} details the proposed framework. Section \ref{sec:chall} discusses the challenges facing an intelligent user interface to enable the framework. Section \ref{sec:conc} concludes with final remarks. 

\section{Related Work} \label{sec:litRev}
Some work about alert and notification systems exists to control notification issuance. Rule-based systems were the most commonly adopted approaches. Consel et al. \cite{consel2015unifying} developed an alerting system for assisted living that reminded an elderly person to take their medication or lock the doors, for example. The system relied on predefined rules to determine when to issue notifications based on their priority. Bazinette et al. \cite{bazinette2001intelligent} allowed users to customize the notifications from a wide range of sources through a single system. \textit{AlertMe} provided users with the capability of customizing alerts and used semantic analysis to issue custom alerts \cite{leonidis2009alertme}. 

More recently, machine learning based approaches have been investigated in restricted domains. Tomavsev et al. \cite{tomavsev2019clinically} developed an approach to issue alerts by predicting whether patients' condition will deteriorate based on test results. Unlike \cite{tomavsev2019clinically}, our work also suppresses or aggregates instantaneous notification of non-urgent alerts. Motivated to prevent missing any notifications, Oh et al. \cite{oh2015intelligent} proposed an intelligent notification system that used user context to provide timely notifications. Finally, Iqbal et al. \cite{iqbal2008effects} proposed a notification management system that determined the best time to interrupt users with notifications by monitoring users' event streams in an application. In intrusion detection, an alert management system that post-processes alerts to determine whether to issue notifications or not has been proposed and aims to minimize the number of missed critical alerts \cite{pietraszek2005data}. 

\section{Background and Definitions}\label{sec:basics}
To better understand the need for our framework and the associated design choices, in this section, we highlight the different complexities and dimensions that our framework needs to take into consideration. We begin by providing a basic understanding of what an event is, something that happens in a dynamic system that is characterized by specific features/conditions. These events can occur at a high or low frequency rate, can be either discrete or continuous with monotonic or oscillatory trajectories. Next, we define alerts and notifications in this framework. 
\begin{itemize}
    \item \textbf{Alert}: A message that is generated by a watcher of a dynamical system when an event occurs.
    \item \textbf{Notification}: A message that is sent to a user when the conditions of an alert are satisfied.
\end{itemize}

Alerts are associated with dimensions such as severity (e.g. error, warning, info, not available), criticality (e.g. critical, non-critical), urgency, and  duration (e.g Repeated vs. one-shot). Notifications are also associated with dimensions such as the notification cycle (periodic vs. aperiodic), and the notifications channels (email, slack, etc.). These dimensions add to the complexity of the problem. To further highlight the complexity, we provide two exemplar scenarios below. 

Consider a non-professional stock market trader who is interested in monitoring the prices of specific mutual funds. She is at work during specific times of the day and cannot be available to check the prices at certain hours and is offline most of the weekdays after 6 pm or is working from home. These conditions infrequently change based on some other personal commitments. 
It would be ideal if a system can learn the user's availability and run queries to issue notifications about prices or events during these times to avoid periodic notifications that will be ignored or having any important news buried in a sea of periodic updates. Furthermore, notifications can also be thought of as potential signals to take actions. For example, in case of money invested in shares, the user can get a notification to either sell her shares because the market has potentially reached a peak point and might fall soon (in this case the alert will depend on the price at which user bought shares and the market's volatility). This action signaling alert can also be sent internally to other subsystems, hence setting off a chained process. 

Consider a travel preapproval process at research institutes that handles employees' travel expenses to conferences. A director could configure the alerting system to notify her of any new or pending travel requests. It would be great if the system intelligently issues or suppress notifications by querying a data store to estimate how many employees will submit a travel request to an upcoming conference. Then, instead of issuing notifications for each submission, it will aggregate them and send a single notification. If the system knows that there are pending requests for an upcoming conference but employees are also submitting requests to another conference at a later date, it will issue a notification about the upcoming conference but suppress all other notifications to prevent the former requests from being overlooked. 

While its possible to handle such scenarios today using rule-based systems, that approach suffers both from an increase in the number of rules as well as the complexity of rules which make them harder to create and maintain. Furthermore, these rules are typically enterprise focused and not user focused. Given the recent advancements in artificial intelligence, we envision a user focused learning framework within a proactive conversational agent. 

\section{Methodology} \label{sec:method}
The framework consists of two main components (Fig. \ref{fig:alert_workflow_detailed}): predictive alert (in blue) and notification (in green) management. A conversational agent with a user interface (UI) allows users to customize their alerts and notifications, through natural language, as well as provide feedback to train the model. 

\subsection{Alert Management System}
The alert management system consists of multiple sub-components. First, an alert customization module takes the user's input and provides recommendations to the user based on the system's behavior. It can be modeled as a classification and clustering problem where similar alerts are clustered and then classified based on relevance to the user. Depending on the amount of available data, more sophisticated learning algorithms can be adopted to achieve more accurate results. Then, an event watcher is configured based on this customization to monitor the system, making use of the system model and predicted behavior to optimization observations. When a change happens in the system to trigger an alert, the alert generation engine creates and forwards it to a classifier that utilizes a taxonomy to characterize the alert. An alert aggregator determines whether to issue or to suppress an alert or to aggregate alerts and issue them at a later date. This information is sent to the notification management system. 


\subsection{Notification Management System}
The notification management system consists of multiple user-centric components. User behavior is modeled to influence the modality, communucation medium (channel), format and time of notifications. Users also have the ability to directly set their preferences using the UI. Notifications are intelligently generated based on the information from the various data sources and components before a notification scheduler determines when to emit the notification, while taking into account the user's schedule.

\subsection{Opportunities for Learning}
\begin{figure*}
    \centering
    \includegraphics[width=0.63\linewidth]{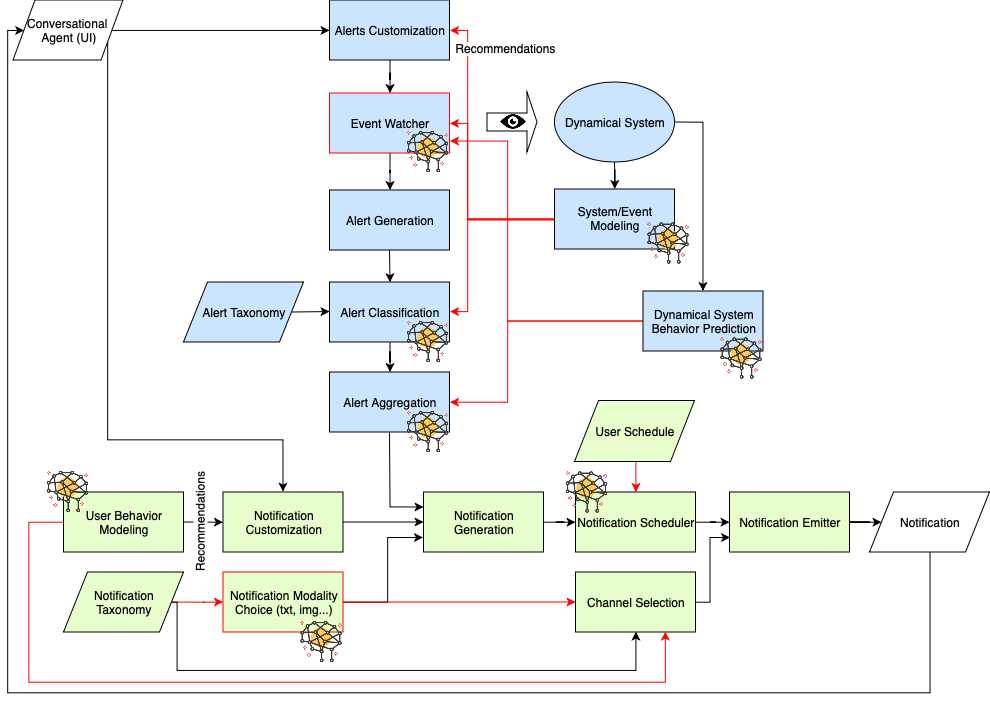}
    \caption{Detailed Workflow of the Alerts Framework}
    \label{fig:alert_workflow_detailed}
\end{figure*}

There are multiple opportunities for learning (depicted by ``brain" clip-arts in Fig. \ref{fig:alert_workflow_detailed}) from users in this framework both in a direct and indirect fashion. At the overall system level, users' direct feedback through the UI can be used as rewards/punishment in a reinforcement learning framework. The system would then learn a model to issue, suppress and aggregate notifications. Indirect feedback can also be used by keeping track of which notifications, issued by the conversational agent, were snoozed, ignored, opened immediately, deleted without being opened, etc. Focusing on individual components of the system, machine learning algorithms such as deep neural networks can be adopted to classify alerts, learn user behavior, determine the notification modality, select the notification channel, and schedule notifications. 

\section{Insights and Challenges} \label{sec:chall}
The proposed framework has multiple advantages. It can intelligently reduce the number of issued notifications thus reducing alert fatigue. Less notifications imply less distractions; therefore, the users' productivity and concentration would increase. Allowing users to provide feedback would enable better customization of the learned model. A byproduct of the system deciding whether to issue or suppress an alert and how to aggregate alerts is implicitly sorting the tasks which improves productivity. Our domain agnostic implementation allows us to easily extend the framework to other domains, especially considering its utility in diverse fields from business to healthcare. Unlike other approaches in the literature, our invention does not adopt a rule-based approach to determine which alerts to issue/suppress, and targets a broader scope than some of the other approaches. Furthermore, it does not require knowledge of the underlying model of the dynamical system it is observing and corresponding alert and notification modules. Learning from unstructured or semi-structured natural language messages generated by the modules reduces the integration overhead. 

The proposed framework faces multiple challenges that must be addressed before real-world deployment. Customizability and generalizability are two conflicting goals. Finding the right tradeoff between both will be crucial to the successful deployment of the framework. Furthermore, classification models make mistakes; and mistakes in one domain can have more severe consequences than in other domains. For example, how can we guarantee that the classification models will not make mistakes in life and death situations? Can we place safeguards around them to ensure that such mistakes are avoided or caught in time? Furthermore, it is difficult to determine why certain machine learning models make the decisions they do. This reduces users' trust in the system and acceptance of it. Therefore, explainability is a crucial component of this system: how can we allow machine learning algorithms to explain their decisions to non-expert users to gain their trust. Another challenge is integrating the system with applications that are generating the events that our system uses to evaluate the alerts and send notifications.

From a user interaction perspective, it is important to design the conversational interface in a user friendly manner to maximize the usefulness of the collected data while minimizing the learning overhead or burden on the user when using the system. The feedback interface should also be simple enough to encourage users to provide meaningful feedback. Clarifying user intents, and enabling disambiguation particularly when the user requests the creation, modification, and deletion of alerts through a conversational interface would be crucial. Another challenge is to empower users to customize notifications, provide feedback, and effectively incorporate that feedback into the reinforcement learning model.

\section{Conclusion} \label{sec:conc}
This work proposes an alerting and notification system that is application-agnostic and personalizable based on predictive and adaptive machine learning algorithms. A method of notifying, escalating, and suppressing alerts should tap into different types of  available datasets to learn user-centric models including: 1) dynamical system behavior, 2) user behavior, and 3) event modeling and classification. This framework can be integrated within proactive chatbots, email services, news services or other services that generate notifications. The system would classify the severity of various alerts, and learn from user data while maintaining confidentiality and privacy of the data. The main challenge, and focus of future work, lies in finding the right data and machine learning models that can learn customizable and generalizable models. 

\bibliographystyle{ACM-Reference-Format}
\bibliography{990_alerts}


\begin{thebibliography}{8}


\ifx \showCODEN    \undefined \def \showCODEN     #1{\unskip}     \fi
\ifx \showDOI      \undefined \def \showDOI       #1{#1}\fi
\ifx \showISBNx    \undefined \def \showISBNx     #1{\unskip}     \fi
\ifx \showISBNxiii \undefined \def \showISBNxiii  #1{\unskip}     \fi
\ifx \showISSN     \undefined \def \showISSN      #1{\unskip}     \fi
\ifx \showLCCN     \undefined \def \showLCCN      #1{\unskip}     \fi
\ifx \shownote     \undefined \def \shownote      #1{#1}          \fi
\ifx \showarticletitle \undefined \def \showarticletitle #1{#1}   \fi
\ifx \showURL      \undefined \def \showURL       {\relax}        \fi
\providecommand\bibfield[2]{#2}
\providecommand\bibinfo[2]{#2}
\providecommand\natexlab[1]{#1}
\providecommand\showeprint[2][]{arXiv:#2}

\bibitem[\protect\citeauthoryear{Bazinette et~al\mbox{.}}{Bazinette
  et~al\mbox{.}}{2001}]%
        {bazinette2001intelligent}
\bibfield{author}{\bibinfo{person}{Vincent Bazinette} {et~al\mbox{.}}}
  \bibinfo{year}{2001}\natexlab{}.
\newblock \showarticletitle{An intelligent notification system}.
\newblock \bibinfo{journal}{\emph{IBM Research Division, Thomas J. Watson
  Research Center, PO Box}}  \bibinfo{volume}{704} (\bibinfo{year}{2001}).
\newblock


\bibitem[\protect\citeauthoryear{Consel, Dupuy, and Sauz{\'e}on}{Consel
  et~al\mbox{.}}{2015}]%
        {consel2015unifying}
\bibfield{author}{\bibinfo{person}{Charles Consel}, \bibinfo{person}{Lucile
  Dupuy}, {and} \bibinfo{person}{H{\'e}l{\`e}ne Sauz{\'e}on}.}
  \bibinfo{year}{2015}\natexlab{}.
\newblock \showarticletitle{A unifying notification system to scale up
  assistive services}. In \bibinfo{booktitle}{\emph{The 17th international
  SIGACCESS conference on computers \& accessibility}}. ACM,
  \bibinfo{pages}{77--87}.
\newblock


\bibitem[\protect\citeauthoryear{Iqbal and Bailey}{Iqbal and Bailey}{2008}]%
        {iqbal2008effects}
\bibfield{author}{\bibinfo{person}{Shamsi~T Iqbal} {and}
  \bibinfo{person}{Brian~P Bailey}.} \bibinfo{year}{2008}\natexlab{}.
\newblock \showarticletitle{Effects of intelligent notification management on
  users and their tasks}. In \bibinfo{booktitle}{\emph{Proceedings of the
  SIGCHI Conference on Human Factors in Computing Systems}}. ACM,
  \bibinfo{pages}{93--102}.
\newblock


\bibitem[\protect\citeauthoryear{Leonidis et~al\mbox{.}}{Leonidis
  et~al\mbox{.}}{2009}]%
        {leonidis2009alertme}
\bibfield{author}{\bibinfo{person}{Asterios Leonidis} {et~al\mbox{.}}}
  \bibinfo{year}{2009}\natexlab{}.
\newblock \showarticletitle{Alertme: A semantics-based context-aware
  notification system}. In \bibinfo{booktitle}{\emph{33rd International
  Computer Software and Applications Conference}}, Vol.~\bibinfo{volume}{2}.
  IEEE.
\newblock


\bibitem[\protect\citeauthoryear{Murphy, Reis, Sittig, and Singh}{Murphy
  et~al\mbox{.}}{2012}]%
        {murphy2012notifications}
\bibfield{author}{\bibinfo{person}{Daniel~R Murphy}, \bibinfo{person}{Brian
  Reis}, \bibinfo{person}{Dean~F Sittig}, {and} \bibinfo{person}{Hardeep
  Singh}.} \bibinfo{year}{2012}\natexlab{}.
\newblock \showarticletitle{Notifications received by primary care
  practitioners in electronic health records: a taxonomy and time analysis}.
\newblock \bibinfo{journal}{\emph{The American journal of medicine}}
  \bibinfo{volume}{125}, \bibinfo{number}{2} (\bibinfo{year}{2012}),
  \bibinfo{pages}{209--e1}.
\newblock


\bibitem[\protect\citeauthoryear{Oh, Jalali, and Jain}{Oh
  et~al\mbox{.}}{2015}]%
        {oh2015intelligent}
\bibfield{author}{\bibinfo{person}{Hyungik Oh}, \bibinfo{person}{Laleh Jalali},
  {and} \bibinfo{person}{Ramesh Jain}.} \bibinfo{year}{2015}\natexlab{}.
\newblock \showarticletitle{An intelligent notification system using context
  from real-time personal activity monitoring}. In
  \bibinfo{booktitle}{\emph{International Conference on Multimedia and Expo}}.
  IEEE, \bibinfo{pages}{1--6}.
\newblock


\bibitem[\protect\citeauthoryear{Pietraszek and Tanner}{Pietraszek and
  Tanner}{2005}]%
        {pietraszek2005data}
\bibfield{author}{\bibinfo{person}{Tadeusz Pietraszek} {and}
  \bibinfo{person}{Axel Tanner}.} \bibinfo{year}{2005}\natexlab{}.
\newblock \showarticletitle{Data mining and machine learning—towards reducing
  false positives in intrusion detection}.
\newblock \bibinfo{journal}{\emph{Information security technical report}}
  \bibinfo{volume}{10}, \bibinfo{number}{3} (\bibinfo{year}{2005}),
  \bibinfo{pages}{169--183}.
\newblock


\bibitem[\protect\citeauthoryear{Toma{\v{s}}ev et~al\mbox{.}}{Toma{\v{s}}ev
  et~al\mbox{.}}{2019}]%
        {tomavsev2019clinically}
\bibfield{author}{\bibinfo{person}{Nenad Toma{\v{s}}ev} {et~al\mbox{.}}}
  \bibinfo{year}{2019}\natexlab{}.
\newblock \showarticletitle{A clinically applicable approach to continuous
  prediction of future acute kidney injury}.
\newblock \bibinfo{journal}{\emph{Nature}} \bibinfo{volume}{572},
  \bibinfo{number}{7767} (\bibinfo{year}{2019}), \bibinfo{pages}{116--119}.
\newblock


\end{thebibliography}

\end{document}